\documentstyle{article}
\begin{document}
\renewcommand{\thesection}{}
\title{Non Singlet StructureFunction At Low x}

\author{D K Choudhury\thanks{Regular Associate, ICTP}
        \\International Center For Theoretical Physics,PO Box586,34100,Trieste,Italy\and Atri Deshamukhya
        \\ Department Of Physics,Gauhati University,Guwahati 781014,India}
%\thanks{Regular Associate ICTP.}
\maketitle
\begin{abstract}We comment on the uniqueness of t-evolution$(t=log(Q^2/\Lambda^2))$ of non-singlet structure functions at low x obtained fromDGLAP equations
\end{abstract}.
\newpage
In recent years,an approximate method of solving DGLAP equations \cite
{kn:ap,kn:alta,kn:gl,kn:dok} at low x has been pursued \cite{kn:cs,kn:scm}. 
In that approach, we expressed the DGLAP equations as partial diffetential 
equations in x (the Bjorken variable)
$x=Q^2/2 p.q$ and $t(t=lnQ^2/\Lambda^2)$ using the Taylor series expansion and 
assuming its validity at low x. One of the limitations of the approach is that 
the solutions reported are not unique\cite{kn:cs,kn:scm}. They are selected as 
the simplest ones with a single boundary condition-the non-perturbative $x$ 
distribution at some initial point $ t=t_{0}$. However, complete solution of 
DGLAP equations with two differential variables in general need two boundary 
conditions\cite{kn:sned,kn:ayre}.

The aim of the present note is to explore the uniqueness of the solution when 
it satisfies some given physically appropriate boundary conditions.
The DGLAP equation for non-singlet structurefunction which evolve independent 
of singlet and gluon distributions\cite{kn:ap,kn:alta,kn:gl,kn:dok} is
\begin{equation}
\frac {\partial F^{NS}}{\partial t}=
[
\frac {A_{f}}{t}
]
[
3 + 4\log(1-x) F^{NS} (x,t) + 2 \int_{x}^{1} \frac {dz}{1-z} (1 + z^2) F^{NS} 
(
\frac {x}{z},t
) 
- 2 F^{NS} (x,t)
]
\end{equation}
where $ t=\log(\frac{Q^2}{\Lambda^2}) $ and $ A_{f}=\frac{4}{33-2 N_{f}},N_{f} $ being the number of quark flavours.
   Let us introduce the variable $ u=1-z $ and note that
\begin {equation}
\frac{x}{1-u}=x\sum_{k=0}^{\infty}u^k 
\end{equation}
   This series (2) is convergent for $  u<1$.Since $ x<z<1 $,so $ 0<u<1-x $ and hence the converegence condition is satisfied.Using (2),we write in (1)
\begin {equation}
F^{NS}(\frac{x}{z},t)=F^{NS}(x,t)+\sum_{l=1}^{\infty}\frac{x^l}{l!}(\sum_{k=1}^{\infty}u^k)^l\frac{\partial^l F^{NS}(x,t)}{\partial x^l}
\end{equation}
which covers the whole range of $ u, 0<u<1-x.$
  Non-singlet structure  function are expected to be well behaved in the entire  $ x$ range,unlike the gluon or singlet structure functions which might diverge for $ x\rightarrow 0$ as in BFKL inspired models\cite {kn:blklf,kn:kms}.It is  therefore justif
ied if the  higher order derivatives $ \frac{\partial^l F^{NS}}{\partial x^l} $ for $ l>1 $ are neglected in (3).This is more justifiable for small $ x( x\ll1)$,yielding
\begin{equation}
F^{NS}(\frac{x}{z},t)=F^{NS}(x,t)+x\sum_{k=1}^{\infty}u^k\frac{\partial F^{NS}(x,t)}{\partial x}.
\end{equation}
Putting (4) in (1) and performing the u-integration,one gets
\begin{equation}
Q(x,t)\frac{\partial F^{NS}(x,t)}{\partial t}+P(x)\frac{\partial F^{NS}(x,t)}{\partial x}=R(x,t,F^{NS})
\end {equation} 
where
\begin{equation}
P(x)=-A_{f}x[2\log(1/x)+(1-x)^2]
\end{equation}

\begin{equation}
Q(x,t)=t
\end{equation}

and 
\begin{equation}
R(x,t,F^{NS})=R^\prime(x)F^{NS}(x,t)
\end{equation}

with
\begin{equation}
R^\prime(x)=-A_{f}[3+4\log(1-x)+(x-1)(x+3)].
\end{equation}

 The general solution of $(5)$ is obtained by solving the following auxiliary systems of  ordinary differential equations
\begin {equation}
\frac{dx}{P(x)}=\frac{dt}{Q(t)}=\frac{dF^{NS}(x,t)}{R(x,t,F^{NS}(x,t)}.
\end{equation}.
If
\begin{equation}
u(x,t,F^{NS})=C_{1}
\end{equation}
and
\begin{equation}
v(x,t,F^{NS})=C_{2}
\end{equation}are two independent solutions   of $ (10)$ , then in general,solution of$ (5)$ is
\begin{equation}
f(u,v)=0
\end{equation}
where $ f$ is an arbitrary function of$u$  and $  v$.

 The auxiliary system $(10)$ has three equations:
\begin{equation}
\frac{dx}{P(x)}=\frac{dt}{Q(t)} .
\end{equation}
\begin{equation}
\frac{dx}{P(x)}=\frac{dF^{NS}}{R(x,t,F^{NS})}
\end{equation}
and
\begin{equation}
\frac{dt}{Q(t)}=\frac{dF^{NS}}{R(x,t,F^{NS})}.
\end{equation}

 Solving $(14)$ and $(15)$ one gets
\begin{equation}
u(x,t,F^{NS})=tX^{NS}(x)
\end{equation}
\begin{equation}
v(x,t,F^{NS})=F^{NS}(x,t)Y^{NS}(x).
\end{equation}
  Solution of $(16)$ is in general not possible. It needs additional informationof  explicit $x$ and $t$ dependence of$ F^{NS}(x,t)$.
 In $(17)$ and $(18)$,$X^{NS}(x)$ and$ Y^{NS}(x)$ are defined by
\begin{equation}
X^{NS}=\exp[-\int\frac{dx}{P(x)}]
\end {equation}
\begin{equation}
Y^{NS}(x)=\exp[\int\frac{dx R\prime(x)}{P(x)}].
\end{equation}
 Explicit analytic forms of  $ X^{NS}(x)$ and$ Y^{NS}(x)$ in the leading $ 1/x $ approximation are
\begin{equation}
X^{NS}(x)\approx\exp[-\frac{1}{2}\log|\log x|]
\end{equation}
while
\begin{equation}
Y^{NS}(x)\approx 1.
\end{equation}
  Note that $(21)$ is strictly valid for
\begin{equation}
x\ll\exp[\frac{-(1-x)^2}{2}].
\end{equation}
 The possible form of $(13)$ linear in $ F^{NS}$ is 
\begin{equation}
u+\alpha v=\beta
\end{equation}
where $\alpha$ and $\beta$ are constants to be determined from physically appropriate boundary  conditions.

 In our earlier communications \cite{kn:cs,kn:scm}we set $\beta $ in$(24)=0$.

 The physically plausible boundary conditions for non-singlet structure functions are
\begin{equation}
F^{NS}(x,t)=F^{NS}(x,t_{0})
\end{equation}
for some low $t=t_{0} $
and
\begin{equation}
F^{NS}(1,t)=0
\end{equation}
for any $ t $.
 While the first one corresponds to a non perturbative input at some low momentum transfer,the second one corresponds to the  expected large$ x$ behaviour of any structure function(singlet as well as nonsinglet) at any scale of momentum transfer consisten
t within  constituent counting rules \cite{kn:rgr,kn:fjy}.
 Using the boundary condition of $ (25)$ and $ (26) $ in $(24)$, we have
\begin{equation}
t_{0}X^{NS}(x)+\alpha F^{NS}(x,t_{0})Y^{NS}(x)=\beta
\end{equation}
and
\begin{equation}
tX^{NS}(1)=\beta
\end{equation}
which leads to
\begin{equation}
F^{NS}(x,t)=(\frac{t}{t_{0}})F^{NS}(x,t_{0})\frac{X^{NS}(1)-X^{NS}(x)}{\frac{t}{t_{0}}X^{NS}(1)-X^{NS}(x)}.
\end{equation}
 As from $(21)$
\begin{equation}
X^{NS}(1)\approx 0,
\end{equation}
$(29)$ yields
 \begin{equation}
F^{NS}(x,t)=(\frac{t}{t_{0}})F^{NS}(x,t_{0})
\end{equation}
which was exactly our previous results\cite{kn:cs,kn:scm}.
 Thus,$ F^{NS}\sim t $ behaviour at low $x$ follows from the two boundary conditions $(25)$ and $(26)$ rather than the adhoc assumption $ \beta=0 $ in $(24)$.
  The auxiliary equation $(14)$ can also have a solution$ \sim u^{-1}(x,t)$ ins
tead of$u(x,t)$ defined in $(17)$.In that case, using the same procedure as earlier, we will have
\begin{equation}
F^{NS}(x,t)=(\frac{t}{t_{0}})F^{NS}(x,t_{0})\frac{X(1)^{-1}-X(x)^{-1}}{\frac{t_{0}}{t}X(1)^{-1}-X(x)^{-1}}
\end{equation}
which is ill defined due to$ (30)$ in $ x\rightarrow 1$ limit and hence excluded on physical grounds.
   
To conclude,we have shown that the  linear rise of  $ F^{NS}(x,t)$ with$ t $ at small $x$ is a unique  prediction of the approximated DGLAP equation with physically plausible boundary conditions $(25)$ and $(26)$.Our result is howevervalid at very small $
x$ when condition $(23)$ is satisfied.For its quantitative estimate,we define a ratio
\begin{equation}
h(x)=\frac{x}{\exp[-\frac{1}{2}(1-x)^2]}.
\end{equation}	
As an illustration,it yields $ h(x)\sim 10^{-3} $ for$ x\sim 6. 10^{-4}$.This small$x$ value is wellwithin the HERA regime \cite{kn:hera} of $x\geq 6. 10^{-6}$,but outside the available CCFR neutrino data \cite{kn:ccfr},where $ x\geq 7.5. 10^{-3}$ 
\section{Acknowledgements}
 This work was completed within the framework of the Associate Scheme of the Abdus Salam International Centre  for Theoretical Physics,Trieste,Italy.Support from DST,India is also acknowledged.
  
\end{document}